\documentclass[preprint2]{emulateapj}
\usepackage{natbib}
\bibpunct[; ]{(}{)}{,}{a}{}{;}

\newcommand{\ccm}{cm$^{-3}$}

\shorttitle{Time-Dependent Density Diagnostics of Solar Flare Plasmas Using SDO/EVE}

\shortauthors{Milligan et al.}

\begin{document}

\title{Time-Dependent Density Diagnostics of Solar Flare Plasmas Using SDO/EVE}

\notetoeditor{the contact email is r.milligan@qub.ac.uk and is the only one which should appear on the journal version}

\author{Ryan O. Milligan\altaffilmark{1}, Michael B. Kennedy\altaffilmark{1}, Mihalis Mathioudakis\altaffilmark{1}, \& Francis P. Keenan\altaffilmark{1}}

\altaffiltext{1}{Astrophysics Research Centre, School of Mathematics \& Physics, Queen's University Belfast, University Road, Belfast, Northern Ireland, BT7 1NN}

\begin{abstract}
\noindent
Temporally-resolved electron density measurements of solar flare plasmas are presented using data from the EUV Variability Experiment (EVE) onboard the Solar Dynamics Observatory (SDO). The EVE spectral range contains emission lines formed between 10$^{4}$--10$^{7}$~K, including transitions from highly ionized iron ($\gtrsim$10~MK). Using three density-sensitive Fe XXI ratios, peak electron densities of 10$^{11.2}$--10$^{12.1}$~\ccm~were found during four X-class flares. While previous measurements of densities at such high temperatures were made at only one point during a flaring event, EVE now allows the temporal evolution of these high-temperature densities to be determined at 10~s cadence. A comparison with GOES data revealed that the peak of the density time profiles for each line ratio correlated well with that of the emission measure time profile for each of the events studied.
\end{abstract}


\section{INTRODUCTION}
\label{intro}

Solar flares are generally considered as increases in the X-ray luminosity on the Sun due to changes in the temperature and density of the coronal plasma. The increase in temperature is readily inferred from the presence of high-temperature ($\gtrsim$10~MK) emission lines in solar flare spectra. However, precise values of the coronal electron density ($N_e$) have been more difficult to ascertain. Quite often these densities are estimated from broadband continuum measurements which can yield values of the flare emission measure ($EM=f\int_V N_e^2 dV$). Deconvolving density values from the emission measure requires a knowledge of the volume of the emitting plasma ($V$, estimated from imaging data) and a possible filling factor ($f$, usually assumed to be unity). Accurate measurements of the electron density are important for understanding both the heating and cooling of flare plasmas, and determining the mechanisms responsible.

A more reliable derivation of electron densities can be made using density-sensitive line ratios, which do not require prior knowledge of the emitting volume nor the filling factor, under the assumption that one of the lines arises from a metastable level. While there have been many studies that have presented the density structure of active regions at coronal temperatures ($\sim$1--2~MK; \citealt{gall01,warr03,mill05}) there have been fewer of the coronal plasma density during flares. \cite{mcke80} and \cite{dosc81} used data from the SOLEX instrument onboard P78-1 to derive time-dependent flare densities from an O VII ratio (2~MK) and found that values reached a maximum of $\sim$10$^{12}$~\ccm~at the peak of the impulsive phase. More recently, \cite{mill11} and \cite{grah11} measured electron densities of 10$^{11}$--10$^{12}$~\ccm~at flare footpoints during their impulsive phases from Fe XII (1.4~MK), Fe XIII (1.6~MK) and Fe XIV (1.8~MK) line ratios using data obtained from the EUV Imaging Spectrometer (EIS; \citealt{culh07}) onboard Hinode. \cite{warr08} identified several Ca XV line pairs (formed at 4~MK) in the EIS wavelength range which are sensitive to densities in the range 10$^{9}$--10$^{11}$~\ccm, while \cite{feld08} identified Ti, Cr, and Mn lines formed above 10~MK which are sensitive to densities from 10$^{10}$--10$^{13}$~\ccm. However, due to the telemetry and planning restrictions implemented on Hinode observations, flare data from these lines are unlikely to become available. 

\begin{figure*}[!t]
\begin{center}
\includegraphics[height=0.8\textwidth,angle=90]{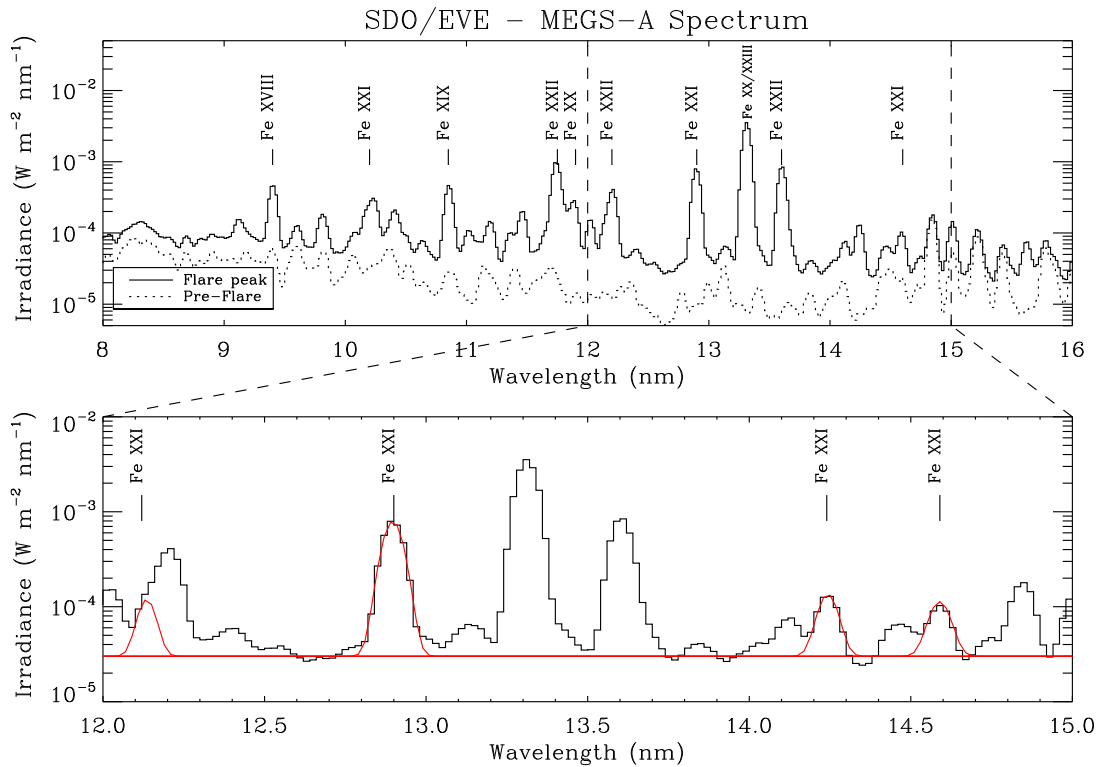}
\caption{Top panel: The 8-16~nm portion of the EVE spectrum showing the presence of high-temperature Fe lines (XVIII-XXIII) at the peak of the X6.9 flare that occurred on 2011 August 9 (solid line), and from a quiescent time before the flare (dotted line). Bottom panel: Expanded view of the 12-15~nm range which contains the four Fe XXI lines used in this study. Overplotted are the fits to each of the lines as well as the background level, indicated by the horizontal line.}
\label{fig:flarelines}
\end{center}
\end{figure*}

Density diagnostics of flaring plasma using emission lines formed above $\sim$10~MK have been even more elusive in recent decades. Among the first to identify such lines were \cite{kast74} using data from the Goddard Space Flight Center (GSFC) scanning spectrograph onboard the Orbiting Solar Observatory (OSO)-5 satellite. These transitions were primarily from Fe XIX--Fe XXIII and found in the 9--14.5~nm portion of the EUV spectrum. Following \cite{dosc73}, \cite{maso79} identified several Fe XXI line pairs in this wavelength range that could serve as density diagnostics during flares for values of $N_e$ between 10$^{11}$--10$^{15}$~\ccm. A preliminary application to OSO-5 data yielded typical flare densities $<$10$^{13}$~\ccm~for temperatures $\sim$10~MK. \cite{maso84} later refined these values, finding 4$\times$10$^{11}$~\ccm~for temperatures between 10$^{6.5}$ and 10$^{7.2}$~K. A similar detailed study of electron densities by \cite{laws84} using the OSO-5 solar flare spectra found inconsistent values of 10$^{12.8}$, 10$^{11.7}$ and 10$^{13.1}$~\ccm~from Fe XX, Fe XXI, and Fe XXII line ratios, respectively. However, the OSO-5 instrument took 2.5~minutes to scan the 9--14.5~nm wavelength range, meaning that each of the lines in a given ratio were recorded at different times, during which the flare was likely to have evolved. The instrument sensitivity as a function of wavelength was also uncertain.

Density diagnostics with improved spectral and temporal resolution were later obtained from the Bent Crystal Spectrometer (BCS) onboard the Solar Maximum Mission. This was a soft X-ray spectrometer that obtained high-resolution spectra in small wavelength intervals near 0.19 and 0.31~nm. An analysis of Fe XX lines in BCS spectra was performed by \cite{phil83} to obtain electron density values. In most cases values of the electron density were found to be $\sim$10$^{11}$~\ccm~and were therefore indistinguishable from the low density limit. However, there has been one report of solar flare densities as high as $10^{13}$~\ccm~from Fe XXII line ratios by \cite{phil96}. In the stellar case, observations made with the Extreme Ultra-Violet Explorer also revealed coronal electron densities as high as 10$^{13}$~\ccm~during the most energetic events \citep{mons96}. Coronal electron densities of this magnitude, if confirmed, could have significant implications for flare physics.

Earlier studies of flare densities using high-temperature line ratio techniques have each focused on a single time interval during the flare, often with an integration time of several minutes. Any time-resolved investigations were undertaken using lines formed at quiescent coronal temperatures. Here we present time profiles of electron density, determined using line ratios with formation temperatures in excess of 10~MK obtained using data from the EUV Variability Experiment (EVE; \citealt{wood10}) instrument onboard the Solar Dynamics Observatory (SDO). Section~\ref{sec:eve_obs} gives an overview of the EVE instrument, the emission lines within the spectral range under consideration and a description of the methods used to analyse the data. In Section~\ref{sec:results} we present results obtained from four X-class flares, while Section~\ref{conc} summarizes the results and conclusions, and discusses the implications.

\section{EVE Observations and Data Analysis}
\label{sec:eve_obs}

EVE acquires full-disk (Sun-as-a-star) EUV spectra every 10~seconds over the 6.5--37~nm wavelength range using its MEGS-A (Multiple EUV Grating Spectrograph) component with a near 100\% duty cycle. The 9--16~nm portion of this wavelength range contains many emission lines from transitions in high-temperature ($\gtrsim$10~MK) Fe ions (XVIII--XXIII), as shown in the top panel of Figure~\ref{fig:flarelines}. These high-temperature iron lines dominate the EVE spectrum during a flare (see \citealt{cham12}).

Although the 9--16~nm range includes several species of high-temperature Fe lines, its coarse resolution ($\sim$0.1~nm) means that many of the observed line profiles were blended with other emission lines, or were too weak to be detected above the level of continuum emission. After investigating all possible density-sensitive line pairs for Fe XIX--XXII listed in \citet[page 171, Figure 6.7]{phil08}, only three pairs of Fe XXI lines were deemed to be reliable (see bottom panel of Figure~\ref{fig:flarelines}). These agree with those identified by \cite{maso79,maso84} and will be discussed in more detail in Section~\ref{fe_xxi_lines}.

The identification of the emission lines present in the EVE spectra was performed by visually comparing the central wavelength of the observed emission features with the CHIANTI (version 7; \citealt{land12}) line list. Synthetic line profiles for this list were generated using the ionization equilibrium files of \cite{brya09}, the flare DEM of \cite{dere79}, coronal abundances \citep{feld92}, initial densities of 10$^{11}$ and 10$^{12}$~\ccm, and at the EVE spectral bin size of 0.02~nm.

\subsection{Fe XXI lines}
\label{fe_xxi_lines}

\begin{figure}[!t]
\begin{center}
\includegraphics[height=0.48\textwidth,angle=90]{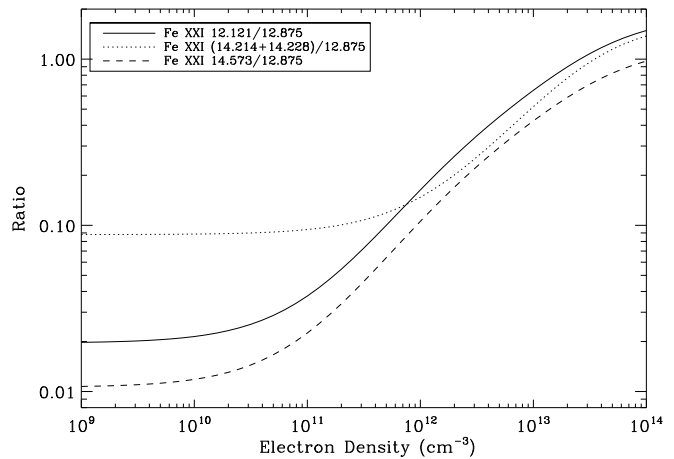}
\caption{Theoretical line ratios for three Fe XXI line pairs calculated using the CHIANTI atomic database.}
\label{fig:density_ratios}
\end{center}
\end{figure}

The strongest Fe XXI emission line in the wavelength range studied is that at 12.875~nm (see also \citealt{maso84}). This is unblended and common to several density-sensitive line pairs, specifically 12.121/12.875, 14.214/12.875, and 14.573/12.875 (see Figure~\ref{fig:density_ratios}). The electron density of the flare plasma can be determined directly from the ratio of the peak intensities. For each density-sensitive line pair, the flux ratio as a function of $N_e$ at the temperature of maximum ionisation fraction, which is $\sim$12~MK for Fe XXI \citep{brya09}, was calculated using CHIANTI. Lines were fitted with a Gaussian function and the peak, rather than integrated, flux used to calculate the ratio values from which the densities were derived. This was to avoid including any potential weaker line emission in the wings of the feature of interest to the total line flux. 

The Fe XXI line at 14.214~nm is blended with another Fe XXI transition at 14.228~nm, and was observed as a single emission feature due to their central wavelengths being separated by only 0.014~nm, which is within one wavelength bin of the EVE spectra. Therefore, the density-sensitive ratio was determined from the sum of the flux of these two lines. This results in the (14.214+14.228)/12.875 ratio being less sensitive to lower densities than if the lines were resolved (dotted line in Figure~\ref{fig:density_ratios}).

The 12.121~nm line was weak in many cases and is blended in the blue wing of Fe XX 12.184~nm but is involved in a ratio which is sensitive to a wide range of densities (solid line in Figure~\ref{fig:density_ratios}). However, measurements of the irradiance were uncertain due to the presence of the stronger Fe XX line except in the largest flaring events. 

The Fe XXI line at 14.573 is blended with a lower temperature line which is present in the pre-flare spectra. By subtracting out a pre-flare profile from the flare spectra effectively eliminates any contribution from the unidentified feature during the SXR peak of the flare, as any enhancement from lower temperature lines are not thought to be significant until later in the flare (for coronal lines) or during the impulsive phase (for chromospheric lines; see \citealt{cham12}).

Each line ratio has a low and high density limit, beyond which the ratio is not suitable for determining electron densities. The low density limit for each of the ratios was taken to be 10$^{11}$~\ccm. 

\subsection{Background Subtraction}
\label{bg_sub}

As some of the Fe lines under investigation are relatively weak, an accurate background subtraction process is crucial for determining the line flux solely due to the flare. \cite{mill12} recently pointed out that the entire MEGS-A wavelength range includes the underlying free-free continuum emission (which is coronal in origin) that becomes enhanced during solar flares, steepening at shorter wavelengths. On top of this lies a pseudo-continuum formed by the close proximity of the many high-temperature Fe lines. However, as the transitions here lie within 2.5~nm of each other, it is reasonable to assume a constant background across all four lines. Therefore, the entire 12-15~nm range was fitted with 17 Gaussian profiles plus a constant background at each time step throughout the flare (see bottom panel of Figure~\ref{fig:flarelines}).

\subsection{Uncertainties}
\label{errors}

In the current release of the EVE data (level 2, version 2), the uncertainties on the irradiance values have not yet been included and the validation efforts are ongoing. As such, each line profile was fit with a Gaussian profile with constant weighting. To calculate uncertainties in the measured densities, the errors in the line intensities (as derived from the least squares fit of the Gaussian profile) were added in quadrature to give upper and lower limits on the flux ratios. There were then converted into upper and lower limits of the density for each time interval. If the lower limit was found to be less than that of the low density limit for a given line pair, then the lower limit was set equal to 10$^{11}$~\ccm. The electron densities corresponding to these limits then provided the range of likely values for each line pair.

\begin{table*}[!t]
\caption{Values and times of peak electron densities and peak emission measures for each of the flares presented in this study.}
\centering
\begin{tabular}{lccccccccc}
\hline
& &  \multicolumn{2}{c}{14.573/12.875} & \multicolumn{2}{c}{(14.214+14.228)/12.875} & \multicolumn{2}{c}{12.121/12.875} & Peak & Time of \\ 
\cline{3-4}
\cline{5-6}
\cline{7-8}
& & Peak & Time of & Peak & Time of & Peak & Time of & GOES EM & peak GOES \\
\multicolumn{1}{c}{Flare} & Date & Log(N$_{e}$) & peak N$_{e}$ & Log(N$_{e}$) & peak N$_{e}$ & Log(N$_{e}$) & peak N$_{e}$ & (10$^{49}$ \ccm)& EM (UT) \\
\hline\hline \\
X6.9 & 2011 Aug 9 & 11.92$^{+0.18}_{-0.26}$ & 08:05:06 UT & 12.09$^{+0.19}_{-0.30}$ & 08:04:46 UT & 12.01$^{+0.30}_{-0.57}$ & 08:04:46 UT & 29.2 & 08:05:20 UT \\[1.5mm]
X5.4 & 2012 Mar 7 & 11.17$^{+0.41}_{-0.17}$ & 00:28:53 UT & 11.52$^{+0.36}_{-0.52}$ & 00:25:43 UT & - &	- & 23.4 &	00:25:43 UT \\[1.5mm]
X2.2 & 2011 Feb 15 & 11.51$^{+0.30}_{-0.87}$ & 01:57:52 UT & 11.69$^{+0.33}_{-0.69}$ & 01:57:42 UT & 11.46$^{+0.30}_{-0.86}$ & 01:57:32 UT & 10.1 & 01:57:42 UT \\[1.5mm]
X2.1 & 2011 Sep 6 & 11.67$^{+0.23}_{-0.40}$ & 22:21:13 UT & 11.89$^{+0.14}_{-0.20}$ & 22:21:33 UT & 11:82$^{+0.51}_{-0.82}$ & 22:21:33 UT & 9.9 & 22:21:07 UT \\[1.5mm]
\hline
\end{tabular}
\label{tab:peak_density}
\end{table*}

\section{Results}
\label{sec:results}

\begin{figure}
\begin{center}
\includegraphics[width=0.48\textwidth]{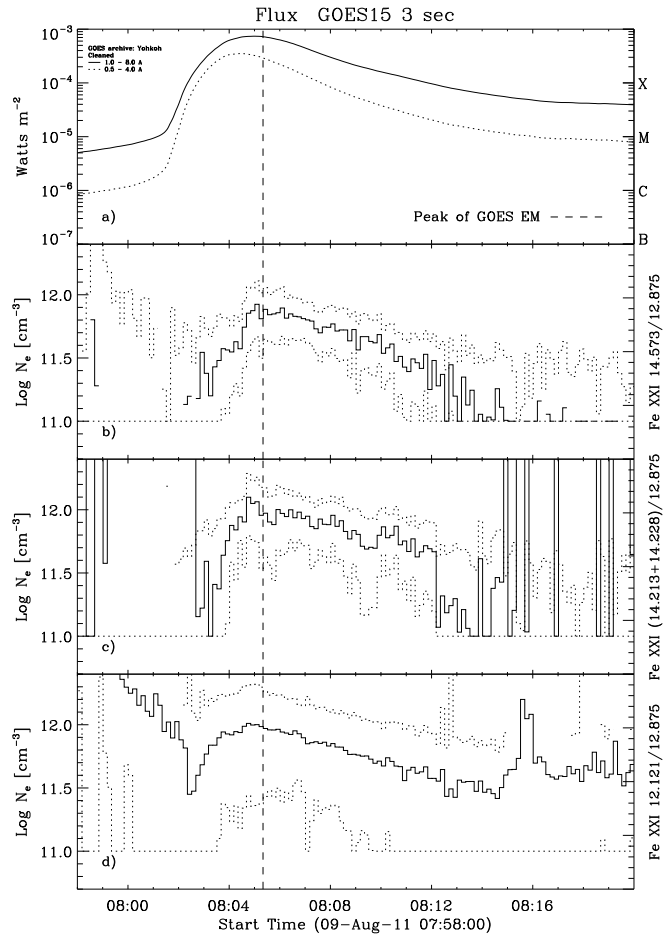}
\caption{Electron density profiles from three Fe XXI line ratios during the X6.9 class flare on 2011 August 9. Panel a shows the GOES lightcurves for the event in the 0.05--0.4 and 0.1--0.8~nm channels. Panels b--d show the time profiles of electron density determined using the 14.573/12.875, (14.214+14.228)/12.875, and 12.121/12.875 Fe XXI ratios, respectively (solid lines). Dotted lines denote the upper and lower density limits. The vertical dashed line in each panel indicates the time of peak emission measure as derived from GOES data.}
\label{f_9aug}
\end{center}
\end{figure}

The methods described in Section~\ref{sec:eve_obs} were applied to four X-class flares observed by EVE during 2011 and 2012, including the X6.9 event that occurred on 2011 August 9 which is the largest flare observed by EVE to date. In the GOES X-ray light curve, emission is observed to increase starting at approximately 07:50~UT, and the peak X-ray flux occurs at 08:05~UT. Electron densities for this flare were derived from the three Fe XXI line ratios described in Section~\ref{fe_xxi_lines} from approximately 08:02~UT until 08:14~UT, during which the values were above the low density limit of the line ratios. The electron density profiles derived from the three line pairs are shown in panels b--d of Figure~\ref{f_9aug} (solid lines), where the dotted lines represent the upper and lower limits of the measured density.

Results from each of the line ratios show a similar trend throughout the main phase of the flare. The electron density increases above the low density limit of the line ratios at approximately 08:02~UT and reaches a peak at around 08:05~UT. There is then a slower decline in density, lasting a period of approximately 9 minutes, at which point the density had decreased to a value below the limit of the line ratios. Each of the three line pairs gave consistent peak electron density values of approximately 10$^{12}$~\ccm. The precise values and times of peak $N_e$ for each line pair is listed in Table~\ref{tab:peak_density} along with their uncertainties. For comparison, the GOES data for each event are also plotted to show the relative timing of the flaring emission (Figure~\ref{f_9aug}a). From the GOES data an emission measure profile can be derived using the methods described in \cite{whit05}. The time of peak electron density is in good agreement with the time of peak emission measure as determined independently from GOES observations (08:05:20~UT, also listed in Table~\ref{tab:peak_density}), as denoted by the vertical dashed line.

This analysis was repeated for three other X-class flares: the X5.4 on 2012 March 7, the X2.2 on 2011 February 15, and X2.1 on 2011 September 6. For all events, consistent peak density values of 10$^{11.5}$-10$^{11.9}$~\ccm~were obtained from each line pair, although reliable values could not be measured for the X5.4 flare using the 12.121/12.875 ratio (see Table~\ref{tab:peak_density}), which may have been due to the dominance of the neighbouring Fe XX line. Despite this limitation, time profiles of 5--30~minutes were obtained for each flare.

\section{Conclusions}
\label{conc}

In this Letter we present techniques for determining time-dependent measurements of the (coronal) electron density using high-temperature ($\sim$12~MK) density-sensitive line ratios during four X-class solar flares. Previous values of electron densities at flare temperatures have only been presented for a small sample of events, at a single time during those flares with integration times of several minutes \citep{phil83,maso84,laws84}. EVE now allows the time profiles of electron densities at flare temperatures to be determined, ultimately, for a statistically significant sample of events throughout the course of solar cycle 24, due to the $\sim$100\% duty cycle of EVE.

After investigating many density sensitive line pairs from lines in Fe XIX--Fe XXII, and accounting for blends within the EVE spectral resolution, the most reliable ratios were found to be Fe XXI 12.121/12.875, (14.214+14.228)/12.875, and 14.573/12.875, in agreement with those identified by \cite{maso84}. Each of these line pairs is sensitive to densities in the range 10$^{11}$--10$^{14}$~\ccm. Applying these diagnostics techniques to four X-class flares revealed peak densities of 10$^{11.2}$--10$^{12.1}$~\ccm, in broad agreement with \cite{maso84}. 

The highest densities measured were for an X6.9 flare and were of the order of 10$^{12}$~\ccm. The fact that consistent values were obtained using each of the line pairs indicate that any potential line blending did not significantly affect the measurements. While comparable values were obtained for the other X-class flares analyzed, no values of electron density above the low density limit could be determined for any M-class flares that were investigated. This limits the suitability of the line ratios investigated to X-class events where the electron density is greater than 10$^{11}$~\ccm, except perhaps in extreme cases. These results tentatively suggest that electron densities are larger in solar flares with a higher peak X-ray flux.

Perhaps more significantly, from the time profiles of the density evolution, the time of maximum density was found to coincide with that of the peak emission measure as determined from GOES data. The GOES 0.05--0.4 and 0.1--0.8~nm passbands contain strong contributions from highly-ionized iron (the Fe XXV/Fe XX complex at 0.185~nm; \citealt{whit05}) as well as free-free continuum emission. It is reasonable to assume, therefore, that the X-ray emission observed by GOES would emanate from approximately the same flaring plasma for which the electron densities were derived.

An accurate determination of the electron density is important for understanding both the heating and cooling of flare plasmas, and the mechanisms responsible. For example, the high densities often inferred from the SXR emitting corona during flares are believed to be a result of chromospheric evaporation, whereby chromospheric material is ablated up into the corona by a beam of nonthermal electrons as inferred from blueshifts of high-temperature emission lines (e.g., \citealt{anto83,czay99,mill09}). Density values are therefore useful for determining the mass rate into (and out of) the overlying loops during evaporation (and condensation). Flare cooling is also believed to be dominated by thermal conduction around the peak of a flare once the injected energy is switched off (e.g., \citealt{raft09}). This transitions to radiative cooling at some point during the decay phase. As this cooling scales as the square of the electron density inaccurate assumptions of $N_e$ may lead to large inaccuracies in the amount of radiative cooling. Similarly, the amount of energy radiated during a flare can be estimated using GOES data. Recently, \cite{ryan12} measured the total radiative losses for $\sim$50,000 flares as a function of GOES class. A constant density of 10$^{10}$~\ccm~was assumed across each event. The results presented here show that large flares can have densities 1--2 orders of magnitude higher, and that that value can vary during individual events. Combining GOES and EVE observations can therefore lead to a more accurate determination of flare energetics.

The volumetric filling factor, $f$, can also be determined by a combination of imaging and spectroscopy once the density is known. EM can be derived from continuum observations from GOES or the Ramaty High-Energy Solar Spectroscopic Imager (RHESSI; \citealt{lin02}), while the volume can be estimated from imaging instruments such as RHESSI, GOES/Soft X-ray Imager, Hinode/X-ray Telescope or the Atmospheric Imaging Assembly (AIA) also on SDO, although AIA is known to saturate during the largest flares. High coronal densities may also help explain the occurrence of coronal HXR emission which is believed to be due to electrons accelerated from a coronal reconnection site impinging upon the underlying loop. Previous estimates place this value as low as 10$^{9}$--10$^{10}$~\ccm, although these were often from much weaker events \citep{kruc08}. Precise knowledge of the flare loop density can be used to establish whether the observed emission is due to thick- or thin-target bremsstrahlung.

Despite being designed to measure changes in the solar EUV irradiance, several recent studies, in addition to the work presented here, have demonstrated how EVE observations can be used to determine some of the fundamental properties of solar flare plasmas: \cite{wood11} presented evidence for a flare `late phase', as well as coronal dimming; \cite{huds11} were able to derive Doppler velocities despite EVE's modest spectral resolution; \cite{mill12} showed that it is possible to determine the timing and energetics of the free-free and free-bound continuum during large events; and \cite{cham12} describes how EVE can be used to measure the thermal evolution of the flaring plasma. Combining these, and other, analyses of solar flares throughout Solar Cycle 24 will give us new insights into their global behaviours and properties.

\acknowledgments
The authors would like to thank the anonymous referee for their comments which helped improve this manuscript. ROM is grateful to the Leverhulme Trust for financial support from grant F/00203/X, and to NASA for LWS/TR\&T grant NNX11AQ53G. He would also like to thank Dr. Phillip Chamberlin for his advice on EVE calibration. MM and FPK acknowledge financial support from the Science and Technology Facilities Council.

\bibliographystyle{apj}
\bibliography{ms}

\end{document}